\documentclass{PoS}

\newcommand{\be}{\begin{equation}}
\newcommand{\ee}{\end{equation}}
\newcommand{\bary}{\begin{eqnarray}}
\newcommand{\eary}{\end{eqnarray}}

\title{Long and short high energy components presented in  GRBs}

\ShortTitle{High Energy emission}

\author{\speaker{N. Fraija}, M. M. Gonzalez, R. Sacahui and W. H. Lee  \\
Instituto de Astronomia, UNAM, Mexico, 04510,  Universidad Nacional Aut\'onoma de M\'exico, 
Circuito Exterior, C.U., A. Postal 70-264, 04510 Mexico D.F., Mexico\\
 \email{nifraija@astro.unam.mx, magda@astro.unam.mx, jsacahui@astro.unam.mx, wlee@astro.unam.mx}}

\abstract{We present a leptonic model on the external shock framework to describe  the long- and short- lasting GeV component of some GRBs. This model was already applied successfully to GRB 090926A, and we extend it to describe the high-energy emission of GRB 090902B and  GRB 090510.   We argue that the high-energy emission consists of two components, one at MeV energies with a duration of a few seconds  during the prompt phase, and a second GeV component  lasting hundred of seconds after the prompt phase. The short high-energy component can be described as SSC emission from a reverse shock and  the longer component arises from SSC emission of the forward shock. The main assumption of our model is that the jet is magnetized and evolves in the thick-shell case. The calculated fluxes and  break energies are all consistent with the observed values}

\FullConference{Gamma-Ray Bursts 2012 Conference -GRB2012,\\
		May 07-11, 2012\\
		Munich, Germany}

\begin{document}


Hadronic \cite{der04, der00} and leptonic \cite{pap96,sar96} models of GRBs have been widely discussed to  explain photons with energies $\geq$ 100 MeV.  In particular, Fraija et al. (2012) and Sacahui  et al. (2012) showed that  the short MeV and long-lasting GeV high-energy components  presented in GRB 980923 and GRB 090926A respectively, could come from  SSC emission in external shocks. We apply the same model in GRB 090902B and  GRB 090510 and show that only the assumption of a magnetized jet is required to explain the complete high-energy emission. Introducing standard values for the input parameters, we obtain break energies, fluxes, duration, etc in agreement with the  observed values. A brief description of the high-energy emission to described for the considered bursts is given below.

GRB 090510 is a short burst observed by both GBM and LAT instruments on board of the FERMI mission\cite{ack10}.  In the first minute post-trigger, 191 events with energy above 100 MeV and 30 events with energy above 1 GeV were detected by LAT. A photon with energy $30.5^{+5.8}_{-2.6}$ GeV was detected by LAT 0.829 s after the GBM trigger and  an extra power law component with power law function of photon index $-1.62\pm 0.03$ was used to describe the high-energy emission. The onset of the high energy spectral component with respect to the beginning of the prompt emission is at $\sim$ 0.1 s.  We are taking into account the photons above 100~MeV starting at $\sim T_{0}+0.8$~s.

GRB 090902B is a long burst observed by GBM and LAT instruments \cite{abd09} at the J2000 coordinates RA=265 $^\circ$  and Dec=27.33$^\circ$. This burst was detected significantly by LAT with 39 photons with energy above 1 GeV.   High energy emission from this burst was detected up to E = $33.4^{+2.7}_{-3.5}$ GeV at 82 s after the GBM trigger. Photons were detected as late as 300 s after the trigger.  The time-integrated spectrum of GRB 090902B is best modeled by a Band function and a power law component with an index of $-2.1\pm 0.1$ and its flux declines as $t^{-1.5\pm0.1}$ over the time interval ($T_0$ + 25, $T_0$+1000 s).  We are taking into account the photons above 100~MeV starting at $\sim T_{0}+13$~s.

GRB 090926A is a long burst observed by Fermi LAT/GBM \cite{ack11} from the J2000 coordinates RA=354.5$^\circ$  and Dec=-64.2$^\circ$. GRB 090926A presents a distinct  high energy power law component \cite{ack11} separate from the known BAND function, fitted  by a Band+CUTPL function with a high energy spectral break at E$_f$=$-1.41^{+0.22}_{-0.42}$ GeV and with a power law index of $-1.72\pm^{+0.10}_{-0.02}$.  A short episode in coincidence with a sharp spike apparent in the light curve above 100~MeV at $\sim T_{0}+10$~s is described  as a power law with a cutoff energy . The flux between 10~keV and 10~GeV for this short episode is $22.29\pm1.60 \times10^{-6}$~erg~cm$^{-2}$~s$^{-1}$ and the power law index is $\lambda = -1.71^{+0.02}_{-0.05}$. 

\section{Leptonic Model (Forward and Reverse shocks) }

In the external shock model, GRB emission is produced when an expanding relativistic shell interacts with  the  circumburst medium producing forward and reverse shocks. In the forward  shock, electrons are accelerated to a power law distribution through the first Fermi mechanism. Thus,  $N(\gamma_e)\,d\gamma_e \propto \gamma_e^{-p}\,d\gamma_e$, with $\gamma_e\geq\gamma_m $ and $\gamma_m=\epsilon_{e,f}(p-2)/ (p-1) m_p/m_e\,\gamma_f$,  where  $\epsilon_{B,f}=  B_f^2/(32\pi\,\gamma^2_f\,\eta_f\,m_p)$   and $\epsilon_{e,f}=  U_e/(4\,\gamma^2_f\,\eta_f\,m_p) $ are  the  magnetic and electron equipartition parameters respectively, $\gamma_f$ is the Lorentz factor of the bulk and $\eta_f$ is the ISM density.   Given the cooling electron Lorentz factor and the deceleration time,  the break energies  of the photons radiated by electrons at a distance $D$ from the source in natural units are given by,

\begin {small}
 \bary\label{synforw}\nonumber
E_{\rm m,f}&\sim& \frac{2^{5/2}\,\pi^{1/2}\, q_e\,m_p^{5/2}\, (p-2)^2}{m_e^3\,(p-1)^2} \, (1+z)^{-1}\,\epsilon_{e,f}^2\,\epsilon^{1/2}_{B,f}\,n^{1/2}_{f}\,\gamma^{4}_{f}\cr
E_{\rm c,f}&\sim& \frac{\pi^{7/6}\, 3^{4/3}\,m_eq_e}{2^{13/6}\,m_p^{5/6}\,\sigma^2_T}\, (1+z)^{-1}\,(1+x_f )^{-2}\,\epsilon^{-3/2}_{B,f}\,n^{-5/6}_{f}\,E^{-2/3}\,\gamma^{4/3}_{f}\cr
\eary
\end{small}

\noindent where $E$ is the isotropic energy. The SSC breaks energies  ($E^{(\rm IC)}_{m,f}\sim\gamma^2_{m},E_{m,f}$ and    $E^{(IC)}_{c,f}\sim\gamma^2_c\,E_{c,f}$)   are also given by\cite{sac12},

\begin {small}
\bary\label{sscf} \nonumber
E^{(IC)}_{\rm m,f}&\sim& \frac{6\,q_e\, m_p^{15/4}}{2^{5/4}\,(3\,\pi)^{1/4}\,m_e^5} \, (1+z)^{5/4}\,\epsilon_{e,f}^{4}\,\epsilon_{B,f}^{1/2}\,n^{-1/4}_{f}\,E^{3/4}\,t_{f}^{-9/4}\cr
E^{(IC)}_{\rm c,f}&\sim& \frac{2^{3/4}\,27\, \pi^{7/4}\,q_e\,m_e^3}{128\,3^{1/4}\,m_p^{9/4}\,\sigma_T^4 } \, (1+z)^{-3/4}\,(1+x_f )^{-4}\,\epsilon_{B,f}^{-7/2}\,n^{-9/4}_{f}\,E^{-5/4}\,t_{f}^{-1/4}\cr
\eary
\end{small}

On the other hand, when  the  reverse shock crosses the shell it heats up and accelerates electrons. Considering the thick shell case, when the ejecta is significantly decelerated, the  synchrotron and SSC break energies are given by \cite{fra12a,fra12b},

\begin {small}
\bary\label{synrev}\nonumber
E_{\rm m,r}&\sim&  \frac{4\,\pi^{1/2}\,q_e\,m_p^{5/2}\,(p-2)^2}{m_e^3\,(p-1)^2}  \,(1+z)^{-1}\,\epsilon_{e,r}^{2}\,\epsilon_{B,r}^{1/2}\,\Gamma^{2}_{r}\,n^{1/2}_{r} \cr
E_{\rm c,r}&\sim&  \frac{9\pi\,2^{1/2}\,m_e\,q_e}{8\,(3^{1/2})\,m_p\,\sigma^2_T}   \,(1+z)^{-1/2}\,(1+x_r+x_r^2)^{-2}\,\epsilon_{B,r}^{-3/2}\,n^{-1}_{r}\,E^{-1/2}\,T_{90}^{-1/2}\cr
E^{(IC)}_{\rm m,r}&\sim&  \frac{2^{21/4}\pi^{3/4}\,m_p^{13/4}\,(p-2)^4}{3^{1/4}\,m_e^5\,(p-1)^4} \,(1+z)^{-7/4}\,\epsilon_{e,r}^{4}\,\epsilon_{B,r}^{1/2}\,\Gamma^{4}_{r}\,n^{3/4}_{r}\,E^{-1/4}\,T_{90}^{3/4}\,,\cr
E^{(IC)}_{\rm c,r}&\sim&  \frac{3^{7/2}\pi\, m_e^3\,q_e}{2^{11}\,m_p^3\,\sigma_T^4   }\,(1+z)^{3/2}\,(1+x+x^2)^{-4}\,\epsilon_{B,r}^{-7/2}\,n^{-3}_{r}\,E^{-1/2}\,\Gamma^{-6}_{r}\,T_{90}^{-5/2}\,,\cr
\eary
\end{small}

\noindent where $T_{90}$ is the burst duration. A detailed description of the model is given in Fraija et al. (2012) and Sacahui  et al. (2012).

\section{Results and Conclusions}

We have used typical \cite{fra12a,fra12b,fra12c} values for $\epsilon_{B,r}\sim 10^{-1}$,  $\epsilon_{B,f}\sim 10^{-4}$,  $\eta_{f}\sim 10^{1}\,cm^{-3}$,
 $\epsilon_{e,r}\sim 0.6$,  $\epsilon_{e,f}\sim 10^{-2}$,  $\gamma_f\sim$ 600 and $\gamma_r\sim$ 1000. The calculated and observed quantities are given in Table 1.
 
 To account for the shorter duration, as compared with the burst duration, of the high-energy emission at energies of hundreds of MeV it is required a value of  $\epsilon_{B,r}\sim 10^{-1}$,  implying a highly magnetized jet.  Moreover, a long-lived keV emission from synchrotron radiation in the forward shock is expected. Some of these bursts present a long-lived tail \cite{gol09} which is interpreted as early afterglow.  In other hand, the duration of the long-lived GeV component is calculated as the time when $E_{f,m}^{IC}$ drops below  $\sim$ GeV, obtaining a value of $\sim 100$~s, in agreement with the observations.

In summary, we have presented  a leptonic model based on external shocks to describe  the long GeV- and short MeV- emission in a unified manner  for GRB 090926A, GRB 090510 and GRB 090902B. The main requirement is a magnetized ejecta. 

\begin{center}\renewcommand{\arraystretch}{0.7}\addtolength{\tabcolsep}{-1pt}
\begin{tabular}{ l c c c c c}
  \hline \hline
\scriptsize{GRBs} & \scriptsize{090510} & \scriptsize{090902B}& \scriptsize{090926A} \\
 \hline 

 {\scriptsize \bf Forward shock}  & & &\\ \hline
{} & {\scriptsize calculated (observed)}       & {\scriptsize calculated (observed)} &  {\scriptsize calculated (observed)} \\





 \scriptsize{$E_{\rm m,f}$ (keV)}   & \scriptsize{330 ( $\sim$ 100 )} & \scriptsize{3.2 ( $\sim$ 10 )}  & \scriptsize{10.13 ( $\sim$ 50 )} \\
 \scriptsize{$E_{\rm c,f}$(eV)}   & \scriptsize{591  ( - )} & \scriptsize{962.5 ( - )} &  \scriptsize{141.7 ( - )} \\
\scriptsize{$E^{(IC)}_{\rm m,f}$(GeV)}  & \scriptsize{2.46 ( $\sim30.5^{+5.8}_{-2.6}$ )} & \scriptsize{16.3  ( $\sim33.4^{+2.7}_{-3.5}$ )} &\scriptsize{9.15 ( $\sim 10$ )} \\
 \scriptsize{$E^{(IC)}_{\rm c,f}(eV)$}  & \scriptsize{$3.8\times10^{-2}$ ( - )} & \scriptsize{$18.25\times10^{3}$ ( - )} & \scriptsize{$18.4$ ( - )}  \\
\scriptsize{Duration of the component (s)}  & \scriptsize{100 ( $\sim 100$ )}& \scriptsize {100 ( $\sim 82$ )}  &\scriptsize{100 ( $\sim 100$ ) }\\  

 \scriptsize{$(\nu F_{\rm \nu max})^{SSC}\,(erg\,cm^{-2}\,s^{-1})$}  & \scriptsize{$3.35\times10^{-5}$ ( $\sim 10^{-5}$ )} & \scriptsize{$2.33\times10^{-6}$ ( $\sim 10^{-6}$ )}  & \scriptsize{$10.9\times10^{-7}$  ( $\sim 10^{-6}$ )}  \\
 
 \hline 
 {\scriptsize \bf Reverse shock}  & & &\\ \hline
{} & {\scriptsize calculated (observed)}       & {\scriptsize calculated (observed)} &  {\scriptsize calculated (observed)}  \\



 \scriptsize{$E_{\rm m,r}$ (keV)}  & \scriptsize{3.03 ( - )} & \scriptsize{165.1 ( - )} & \scriptsize{0.17 ( -  )}  \\
 \scriptsize{$E_{\rm c,r}$(eV)}  & \scriptsize{$1.8\times10^{-3}$ ( - )} & \scriptsize{$1.1\times10^{-3}$ ( - )}  &  \scriptsize{$1.1\times10^{-3}$ ( - )}  \\
 \scriptsize{$E^{(IC)}_{\rm m,r}$(MeV)}  & \scriptsize{$30.0\times10^3$ ( $10^3$)} & \scriptsize{414.3 ( $10^3$  )}  &\scriptsize{414.3 ( 400 ) }  \\
 \scriptsize{$E^{(IC)}_{\rm c,r}(eV)$}& \scriptsize{$6.34\times10^{-7}$ ( - )} & \scriptsize{$0.7 \times 10^{-5}$ ( - )}  & \scriptsize{$0.7\times10^{-5}$ ( - )}   \\
\scriptsize{Duration of the component (s)} & \scriptsize{6  ( $< 1$ ) }& \scriptsize {5 ( $\sim 4$ ) }  &\scriptsize{5 ( $\sim 1$ )  } \\  
 \scriptsize{$(\nu F_{\rm \nu max})^{SSC}\,(erg\,cm^{-2}\,s^{-1})$}  & \scriptsize{$1.4 \times 10^{-6}$  ( $\sim 10^{-6}$ )} & \scriptsize{$8.47 \times 10^{-7}$ ( $\sim 10^{-6}$ ) }  & \scriptsize{$8.2 \times 10^{-6}$ ( $\sim 10^{-6}$ )}  \\
 \hline


\end{tabular}
\end{center}

\begin{center}
\scriptsize{\textbf{Table 1. Calculated quantities are given. For comparison, the corresponding observed values are given when available. }}\\
\scriptsize{}
\end{center}

\end{document}